# On Some Peculiarities of Dynamic Switch between Component Implementations in an Autonomic Computing System

*Igor Mackarov*

*Behavior of the delta algorithm of autonomic switch between two component implementations is considered on several examples of a client-server systems involving, in particular, periodic change of intensities of requests for the component. It is shown that in the cases of some specific combinations of elementary requests costs, the number of clients in the system, the number of requests per unit of time, and the cost of switch between the implementations, the algorithm may reveal behavior that is rather far from the desired. A sufficient criterion of a success of the algorithm is proposed based on the analysis of the accumulated implementations costs difference as a function of time. Suggestions are pointed out of practical evaluation of the algorithm functioning regarding the observations made in this paper.*

**Introduction.**

One of important modern issues of software engineering is development and description of distributed, pervasive computing systems consisting of various, heterogeneous resources, critical to the computer power usage, enabling reconfiguration and scaling the environment [1]. In a number of special cases reconfiguration must be *dynamic*.

The idea of dynamic reconfiguration of the systems gave rise to the concept *of autonomic computing* letting computing systems manage themselves (like autonomic nervous system governs the human body parameters) depending of high-level objectives from administrators. Autonomic systems should involve such fundamental principles as self-configuration, self-optimization, self-healing, self-protection. The principal architectural structure of an autonomic element involves an autonomic manager responsible for communication of the element with the outer part of the general system, and a set of *autonomic components* coordinated by the autonomic manager by means of persisting policies, facts, and rules [2-3].

The success of an autonomic system behavior is essentially determined by ability to detect or predict overall performance that is actually the ground for management of autonomic components, in particular, for activation of an appropriate component implementation. For this, establishing of mathematical abstractions and models giving criteria governing the sequence of switches between component implementations is an important point of autonomic computing [2-5].

**The delta algorithm**

The delta algorithm of the switch between two component implementations, described and studied in [6] can be considered as one of such models. The key abstractions of the algorithm are the costs of both implementations (they may be treated as time intervals needed to send a request for the implementation and to obtain a response), and costs of switches between them (time needed to handle the switch request).

One of the most important moments pointed out as open issues in [6] is the question about work of this algorithm in concrete conditions. The present paper analyses the algorithm behavior for several important classes of the sequences of requests for component implementations.

Before this analysis, following [6], we are going to describe the way the algorithm chooses the moment to switch between the implementations in two forms.

**Mathematical description.**

If we have two implementations $I_i, i \in \{1,2\}$ and designate the non-active one $\overline{I_i}$, and if for the sequence of $k$ requests $r_i$, $i = 0,..k$ there is $j$, $j \leq k$, such that for the total costs of both implementations accumulated during the $j - k$ subsequence of these requests

$$Cost((r_j,...,r_k), \overline{I_i}) \leq Cost((r_j,...,r_k), I_i) - SC$$

where $SC = SC_1 + SC_2$ is the so-called *round trip switching cost*, i.e., the sum of costs of switching from one implementation to the other and vice versa, then the algorithm makes a decision to switch to a non-active implementation.

**Description in terms of a program code.**

This idea may be implemented in terms of Java or C++ code as follows (cf. [6]):

```
impl1Cost = 0;
impl2Cost = 0;
minDelta = 0;
timeToSwitch = false;
while (!timeToSwitch)
{
      impl1Cost += Cost(r, impl1);
      impl2Cost += Cost(r, impl2);
      temp = impl1Cost - impl2Cost;
      minDelta = min(minDelta, temp);
      if(impl1Cost - implCost2 - minDelta >= SC)
        timeToSwitch = true;

}
// making switch from the first
// to the second implementation
```

The code for the *second-first* implementation switch is quit analogous.

**The distributed pub/sub problem.**

Also concretize implementations for the so-called *pub/sub model*, introduced in [6]. This model involves a server holding a database and one or more clients that can read from the database and update it. Two implementations are just *nonsubscription mode* when any read or write request is being directed to the server database, and *subscription mode* when each client has a local copy of the database residing on the server, with the data being read from the local database, and elementary writing consisting in a request to the server to update its database and in notification from the server to all the clients (except the one that writes) to update their local database images.

Again following [6], it is quit natural to assume the elementary read and write operations costs to be as follows:

- *Nonsubscription mode:*

    $C_{nw} = 1$

    $C_{nr} = 2$

- *Subscription mode:*

    $C_{sw} = 2 + (N_{cl} - 1)$     ($N_{cl}$ is the total number of participating clients)

    $C_{sr} = 0$                    (assume that reading from the client local database image costs negligibly little).

Also it is natural to suppose that the cost of switch from *nonsub* to *sub* mode is equal to the cost of sum of reading requests covering all the database because actually what we need for this switching is reading the whole of the database from the server to form its local image:

$$SC_{ns} = \sum_{i=0}^{\substack{number\ of\ items \\ in\ database}} r_i \qquad (1)$$

Besides, evidently $SC_{ns} \gg SC_{sn}$ and therefore

$$SC \approx SC_{ns}$$

To investigate the delta algorithm work with this model, let us make a supposition that the number of requests to the implementations is big enough so we can describe the

"intensity" of the requests, i. e., the number of requests per unit of time, by sufficiently "smooth" mathematical functions (we will concretize the requirement to these functions later). Now let the numbers of read and write requests per unit of time are given by formulae:

$$n_r = \frac{1}{2}W(1-\cos(\omega t)), \qquad n_w = \frac{1}{2}W(1+\cos(\omega t)) \qquad (2)$$

That may take place, for instance, in the case of some distributed system for accumulation and processing of information from a scientific experiment. We may suppose that at the beginning of a cycle (which may be a working day or the some stage of an experimental program) writing requests dominate (active accumulation of experimental data obtained), whereas closer to the end of the cosine's semi-cycle the processing and treatment of the data obtained by the system clients begin to prevail, so intensity of reading reaches its maximum $W$ (the time length of the cycle can be regulated by phase factor $\omega$).

The increments to the costs of *nonsub* and *sub* modes are therefore

$$\begin{aligned} I_n &= C_{nr}n_r + C_{nw}n_w \\ I_s &= C_{sr}n_r + C_{sw}n_w \end{aligned} \qquad (3)$$

Assuming $N_{cl} = 2$ (some small model system consisting of a server and two clients), we can find the difference between *accumulated* costs of nonsub and sub modes:

$$\Delta_2 = \int_0^t (I_n - I_s)\,dt = -2\frac{W}{\omega}\sin(\omega t) \tag{4}$$

Now let $\lambda$ be the part of database size (items) that is covered by the read requests during one period of functions (2). Then, according to our suppositions about the switch cost and great number of requests per unit of time, we are passing from summation in (1) to integration:

$$SC \approx SC_{ns} = \frac{1}{\lambda}\int_0^{2\pi/w} C_{nr} n_r\,dt = 2\pi\frac{W}{\omega\lambda} \tag{5}$$

It can be easily seen from (4)-(5) that if the database size is such that the whole of the readings throughout the cycle covers $\pi/2 \approx 1.6$ portion of the database then *SC* is exactly equal to double amplitude of the accumulated costs difference, i. e., the distance between horizontal lines on Fig. 1.

Looking at Fig.1, suppose now that at the initial moment of time nonsub implementation is active. Then the cost difference function monotonically decreases up to point A, which means that the nonsub active implementation is cheaper and we have an optimal situation. But after passing point A the *sub* implementation gets cheaper since the curve rises. But the delta algorithm will evidently decide to switch to the sub implementation only upon the curve's coming to point B (the difference between accumulated costs became equal to SC, that is the distance between the curve's maximum and minimum).

What happens then is quit evident. Our function decreases again; we have more expansive *sub* implementation active, and cheaper *nonsub* implementation passive. But again the delta algorithm can figure out this only after the curve reaches point C. Then, apparently, the situation repeats. So the algorithm chooses wrong implementation every time even though we started with correct implementation!

This is, however, an extreme case. If *SC* has smaller value than the one considered above, i. e.,

$$0 < SC < 2W$$

then for the accumulated cost we have

$$C = \int_0^{\pi/2\omega} I_n dt + \int_{\pi/2\omega}^{(\pi+\arcsin(SC-1))/\omega} I_n dt + \int_{(\pi+\arcsin(SC-1))/\omega}^{3\pi/2\omega} I_s dt + \int_{3\pi/2\omega}^{(2\pi+\arcsin(SC-1))/\omega} I_s dt + \int_{(2\pi+\arcsin(SC-1))/\omega}^{5\pi/2\omega} I_n dt = \frac{15}{4}\pi - \frac{9}{2} + 4SC \quad (6)$$

$$\frac{15}{4}\pi - \frac{9}{2} < C < \frac{15}{4}\pi + \frac{7}{2}$$

The left limit corresponds to ideal behavior when a cheaper implementation is always active (in this case the first, the third, and the fourth terms in (6) are not equal to zero, so the falling parts of the Fig. 3 graph involve *nonsub* mode, the rising parts have the *sub* mode active). The right limit provides the worst opportunity when we always have a more expensive implementation active (the first, the second, the fourth terms are not zeros, the falling graph parts have *nonsub* mode, the rising parts have the *sub* mode active). This is actually the situation examined above.

We can observe the same behavior of the delta algorithm, for instance, in the case of *3 clients* (see Fig. 2). In this case, evidently, the accumulated costs difference

$$\Delta_3 = -\frac{2W}{\omega}\sin(\omega t)(1+\cos(\omega t))$$

and in the case when the read requests cover the database $\frac{8}{9}\pi\sqrt{3} \approx 4.8$ times during one period of functions (2) the round trip switching cost *SC* again equals double amplitude of the accumulated costs difference and the situation with wrong choosing of the components actually repeats.

Apparently, for such "wrong algorithm's behavior", the accumulated costs difference is to be a purely periodic function. Besides, some "coherence" between elementary costs, number of clients, number of requests per unit of time, and the switch cost is needed. This does not necessarily occur for each instance of the system considered. For example, in the case of *ten clients* the accumulated costs difference is

$$\Delta_{10} = \frac{1}{2}\frac{W}{\omega}(2\cos(\omega t) - 11\sin(\omega t) - 9t\omega - 2)$$

(see Fig. 3). This is "almost" monotonic function and evidently the algorithm either won't make a switch at all (if we start from the cheapest implementation) or will switch to the cheapest implementation once as soon as the function reaches the value –*SC*. In both cases we will obtain ideal or close to ideal behavior.

Summarizing this part of the delta algorithm behavior observations, we can say that in an important particular case of cyclic intensities of requests to the component this algorithm often reveals behavior rather far from the desired, so further improvement of the forecast method or accurate clarification of cases when the algorithm can or cannot be applied may be needed here.

**On conditions of correct work of the algorithm.**

Establish a certain class of time dependencies of the implementations accumulated costs difference with which the algorithm involved will work correctly. First, it should be pointed out that, since the function of accumulated costs difference is a by-linear form (3) of requests intensities and elementary requests costs, we could think of its *second derivative* (determining the function plot curvature) as the rate of the changeability of the requests per unit of time, or intensity of change of the requests costs which may depend on the network stability in the case of a distributed system with remote components.

So we can consider the involved function second derivative absolute value as some basic quantity that can be estimated either by direct analysis of the system performance, or by numerical processing (interpolation) of the curves obtained from previous system sessions.

If the implementations accumulated costs difference function, whose second *continuous* derivative is less or equals some value $d_2$, has at some initial consideration point $t=0$ first derivative $d_1>0$, it's curve will be located *over* the parabola whose derivatives at the initial point are also $d_1$, $d_2$ at $t>0$:

$$p(t) = \frac{d_2}{2}t^2 + d_1 t$$

Evidently, such a parabola reaches its *maximum* (in this case $d_2<0$), which is greater or equals 2*SC*, if

$$|d_1| > 2\sqrt{SC \cdot |d_2|} \qquad (7)$$

Apparently, this is a criterion of the considered difference function monotonic increase by a value *2SC*. During this growth, if the system's cheaper implementation is not active the algorithm instructs the application to switch to it as soon as the function increases by *SC*. Then another increase by *SC* follows, so the algorithm's decision is correct because the newly activated implementation turned out to be cheaper by at least the switch cost! Had we a cheaper implementation active at the initial moment, we would observe the optimal performance without the switch.

It the case $d_1<0$ at the initial consideration point, our curve will be located below the parabola having *minimum*, $d_2$ being more than 0. So monotonic *decrease* of the function involved by a value $2SC$ will be provided and we can observe the same result as in the previous paragraph.

**Conclusion and open issues**

We may suppose now that taking into consideration criterion (7) *using* the delta algorithm, may increase the overall efficiency of its work. It might be even more important to emphasize that both diminishing of medium absolute value *second derivative* of the implementations accumulated costs difference function (by increase of the network connection stability and growth of the requests restructuring characteristic time) and increase of the absolute medium value of *first derivative* (via growth of the mean requests activity) can raise the overall efficiency of the delta algorithm performance on any system using dynamic switch between implementations.

Finally it should be mentioned that since real distributed computer systems are often very complicated and varied it may be of great interest and importance to observe the efficiency of the criterion (7) in concrete cases as well as real systems performance in the case of cyclic requests activity considered in the first part of the paper.

**Igor Mackarov.** *Maharishi University of Management, SU 221, 1000 North 4$^{th}$ Street, Fairfield, Iowa 52557* (**imackarov@mum.edu**)**.** I. Mackarov graduated from Moscow Institute of Physics and Technology in 1988 with the degree of engineer-physicist. In

1988-1991 had a post-graduate study at Institute for Problems in Mechanics of Russian Academy of Sciences (Moscow). In 1992-2001 did research on Computational Fluid Dynamics in scientific and technological institutes in Belarus. In 2001-2003 worked in International Business Alliance (Minsk, Belarus), an official partner of IBM. In 2003 entered MS Computer Professionals program at Maharishi University of Management (Fairfield, Iowa). Published nine articles on numerical and analytical studies of different problems of Continuous Media Mechanics, numerical simulation of technological mechanical processes.

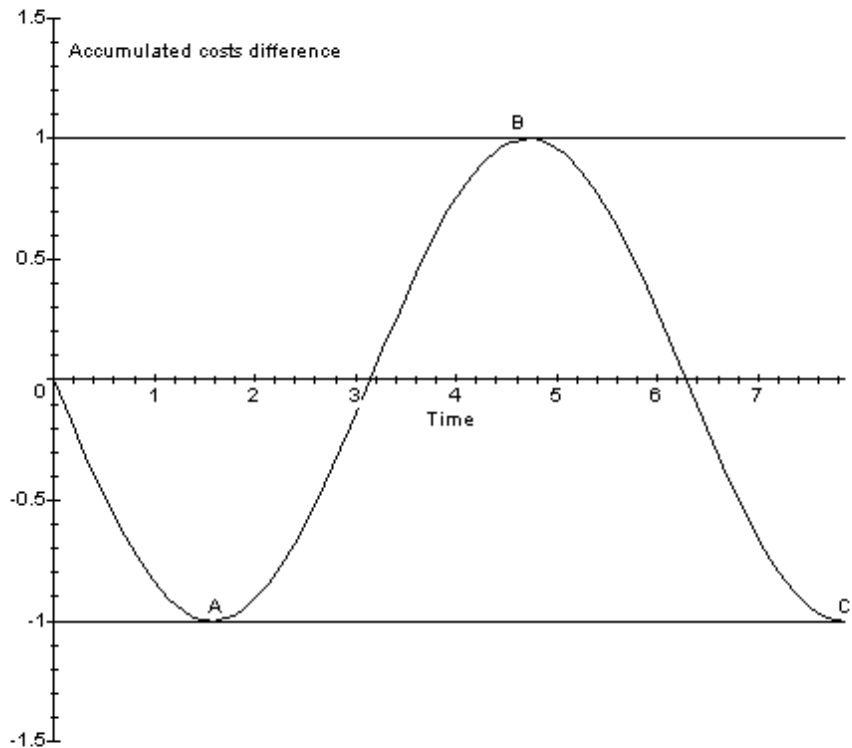

**Figure 1.** Difference between accumulated costs of *nonsub* and *sub* implementations for the system consisting of a server and two clients as a function of time. Both time and costs here are defined to be dimensionless in the scale of $\omega$ and $W$ respectively.

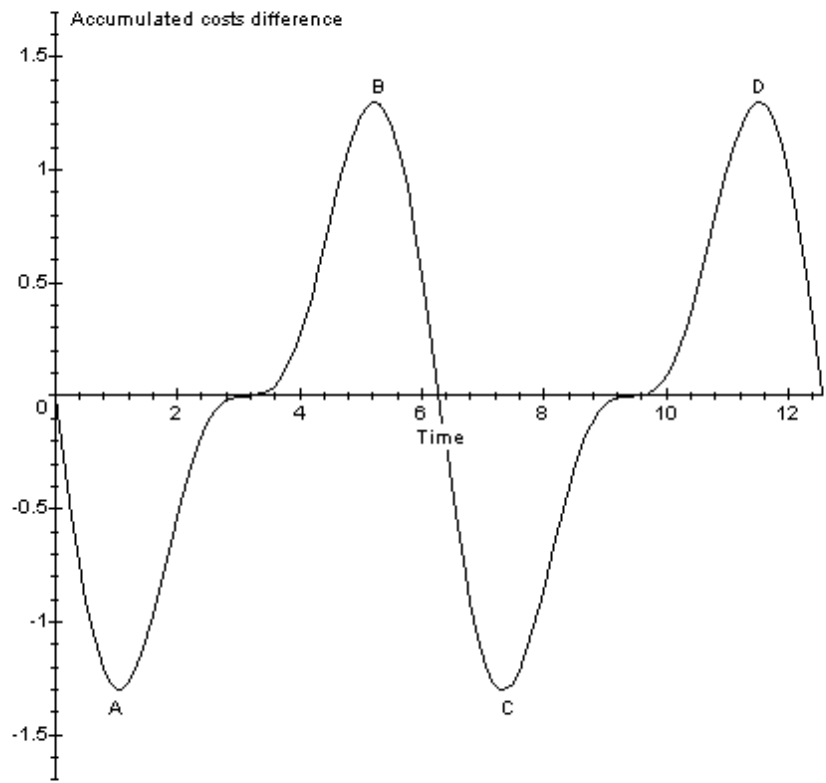

**Figure 2.** Difference between accumulated costs of *nonsub* and *sub* implementations for the system server - three clients as a function of time. Both time and costs here defined to be dimensionless in the scale of $\omega$ and *W* respectively.

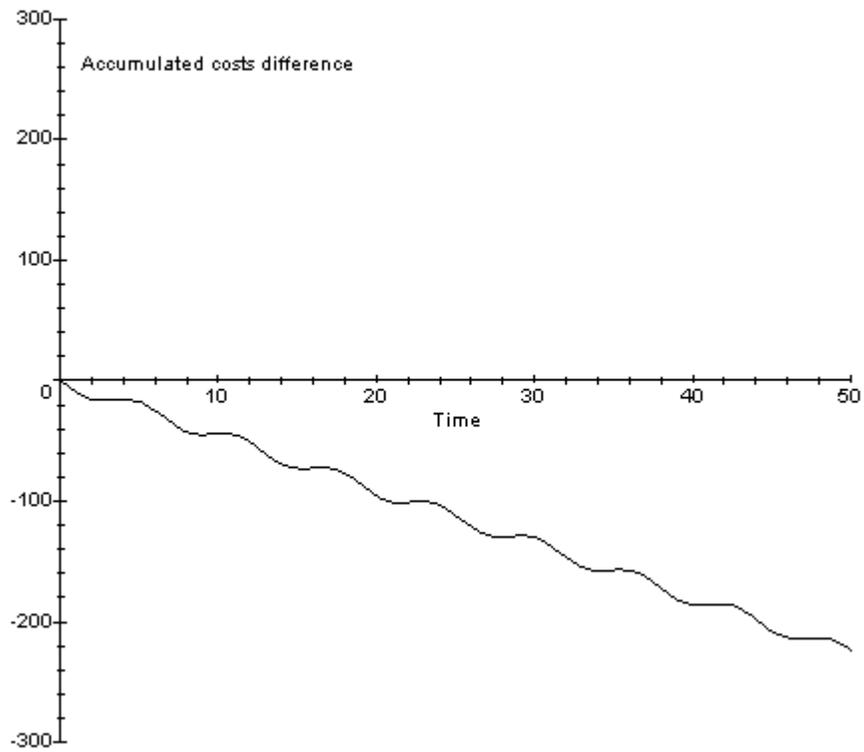

**Figure 3.** Difference between accumulated costs of nonsub and sub implementations for the system server - ten clients as a function of time. Both time and costs here defined to be dimensionless in the scale of $\omega$ and W respectively.